\documentclass[conference]{IEEEtran}

\AtBeginDocument{%
  \providecommand\BibTeX{{%
    \normalfont B\kern-0.5em{\scshape i\kern-0.25em b}\kern-0.8em\TeX}}}

\usepackage{graphicx}
\usepackage{cite}

\begin{document}

\title{Data Driven Modeling Social Media Influence using Differential Equations}

\author{\IEEEauthorblockN{Bailu Jin}
\IEEEauthorblockA{Cranfield University\\
Address: College Rd, Cranfield,Bedford, UK\\
Email: bailu.jin@cranfield.ac.uk }
\and
\IEEEauthorblockN{Weisi Guo}
\IEEEauthorblockA{Cranfield University\\
Address: College Rd, Cranfield,Bedford, UK\\
Email: Weisi.Guo@cranfield.ac.uk}
}

\maketitle

% \IEEEtitleabstractindextext{%
\begin{abstract}
Individuals modify their opinions towards a topic based on their social interactions. 
% Empirical research in psychology has shown that individuals influence each other by seeking similarity or conforming under social pressure. 
Opinion evolution models conceptualize the change of opinion as a uni-dimensional continuum, and the effect of influence is built by the group size, the network structures, or the relations among opinions within the group. However, how to model the personal opinion evolution process under the effect of the online social influence as a function remains unclear. Here, we show that the uni-dimensional continuous user opinions can be represented by compressed high-dimensional word embeddings, and its evolution can be accurately modelled by an ordinary differential equation (ODE) that reflects the social network influencer interactions. 
% Our three major contributions are: (1) introduce a data-driven pipeline representing the personal evolution of opinions with a time kernel, (2) based on previous psychology models, we model the opinion evolution process as a function of online social influence using an ordinary differential equation, and (3) applied Our opinion evolution model to the real-time Twitter data. 
We perform our analysis on 87 active users with corresponding influencers on the COVID-19 topic from 2020 to 2022. The regression results demonstrate that 99\% of the variation in the quantified opinions can be explained by the way we model the connected opinions from their influencers. Our research on the COVID-19 topic and for the account analysed shows that social media users primarily shift their opinion based on influencers they follow (e.g., model explains for 99\% variation) and self-evolution of opinion over a long time scale is limited.
\end{abstract}

\section{Introduction}

\subsection{Background}
Since early research stretching back to the 1940s, social influence is proved to have a vital effect on opinion modification. Several empirical research in psychology has shown that individuals evolve their opinions towards a topic since they seek similarity or conform under social pressure. Sociologists modelled the social influence as a force, determined by the size, the network structure, and the relations among opinions, to mathematically capture the observed personal opinion evolution phenomenon. 

In an age of internet-based tools becoming one of the primary sources of communication repertoires, online interaction and interpersonal communication are rapidly converging \cite{flanagin2017online}. Recently we have used the social influence theories to analyse online social networks (OSN). The online social influence can involve and interact with real-world crises, such as the Russian-Ukraine conflict \cite{conflict19}. Therefore, understanding the mechanism of online social influence \cite{influencenetwork19, biran-etal-2012-detecting} is critical.

However, how to model the personal opinion evolution process under the effect of the online social influence as a function remains unclear. In this preliminary paper, we aim to apply the social influence modelling method to the online social network and evaluate the online opinion evolution model using real-time Twitter data. 

\subsection{Related Work}

\subsubsection{Empirical research in psychology}
Individuals evolve their opinions, attitudes, or stances towards topics through their social interactions. Several empirical research in psychology has studied the phenomenon of opinion evolution during interpersonal interaction. Back in 1995, Asch developed one empirical experiment on social conformity, which has shown that people would modify their opinions to seek similarity with others in the group \cite{asch1956studies}. Other experiments on small group behavior, decision making, and innovation diffusion showed how interactions reduce opinion differences between person.
% \cite{sherif1979research,myers1982polarizing,rogers1995diffusion}.
In 2012, a 61-million-user experiment was launched on Facebook during the 2010 US congressional elections \cite{bond201261}. The results show that human behavior is also amendable through interventions from online networks. 

\subsubsection{Models of Social Influence}
Based on empirical conclusions, social influence modellings are proposed to explain the social phenomena of opinion clustering or opinion controversy. French in \cite{french1956formal} introduced the earliest formal model on the opinion evolution in a group. 

Starting from the formal model, the change of opinion is always conceptualized as a uni-dimensional continuum and determined by the size, the network structures, or the relations among opinions within the group. French-DeGroot model showed how social influence always leads to opinion consensus using the assumption that people will always influence each other in the group \cite{degroot1974reaching}. However, opinion consensus is not the only outcome from group discussion experiments. To explain the opinion clustering, the Hegselmann-Krause model adds a bounded confidence attribute to block the influence from opposite opinions \cite{hegselmann2002opinion}. 

\subsection{Contributions \& Novelty}

Several studies investigate the opinion evolution phenomenon on a topic in Twitter. However, most studies focus on the sentiment evolution of the majority of users.
% \cite{lwin2022evolution, SCC16}.
In this paper, we propose the analysis of opinion evolution on a personal level. This paper's three major contributions are: 
\begin{itemize}
    \item We introduce a data-driven pipeline representing the personal evolution of opinions with a time kernel.
    \item Based on previous psychology models, we model the opinion evolution process as a function of online social influence using an ordinary differential equation. 
    \item Our opinion evolution model is applied to the real-time Twitter data under the COVID-19 topic. We find that social media users primarily shift their opinion based on influencers they follow, and self-evolution of opinion over a long time scale is limited.
\end{itemize}
% The remainder of this paper is organized as follows. In Section 2, we formulate our opinion evolution model. Section 3 presents our pipeline on filtering available users and pre-processing tweet contents with word embedding and dimensionality reduction. In Section 4, we propose our regression result and the analysis. Section 5 concludes this paper and the future work.

\section{Social Network Opinion Differential Equation Model}

% \subsection{Model Introduction: Uni-dimensional Opinion Evolution}

% In this part, we present the reader the way we infer influence and opinion to regress our opinion model using online social media data.

% The previously introduced social influence model mainly conceptualizes the opinion using pro-event and post-event psychology survey questions. In online social networks, we use the text content posted by one individual to represent the individual's opinion. We aim to apply the social influence modelling method to the online social network, so we use the compressed word-embedding vectors to capture the vibration of opinion. Our model is based on the fact that group size, network structures, or relations among opinions determine the influence effect, and the opinion evolution mainly develops due to the impact. 
% \subsection{Data Driven Modeling}
\begin{figure}[t]
  \centering
  \includegraphics[width=3.3in]{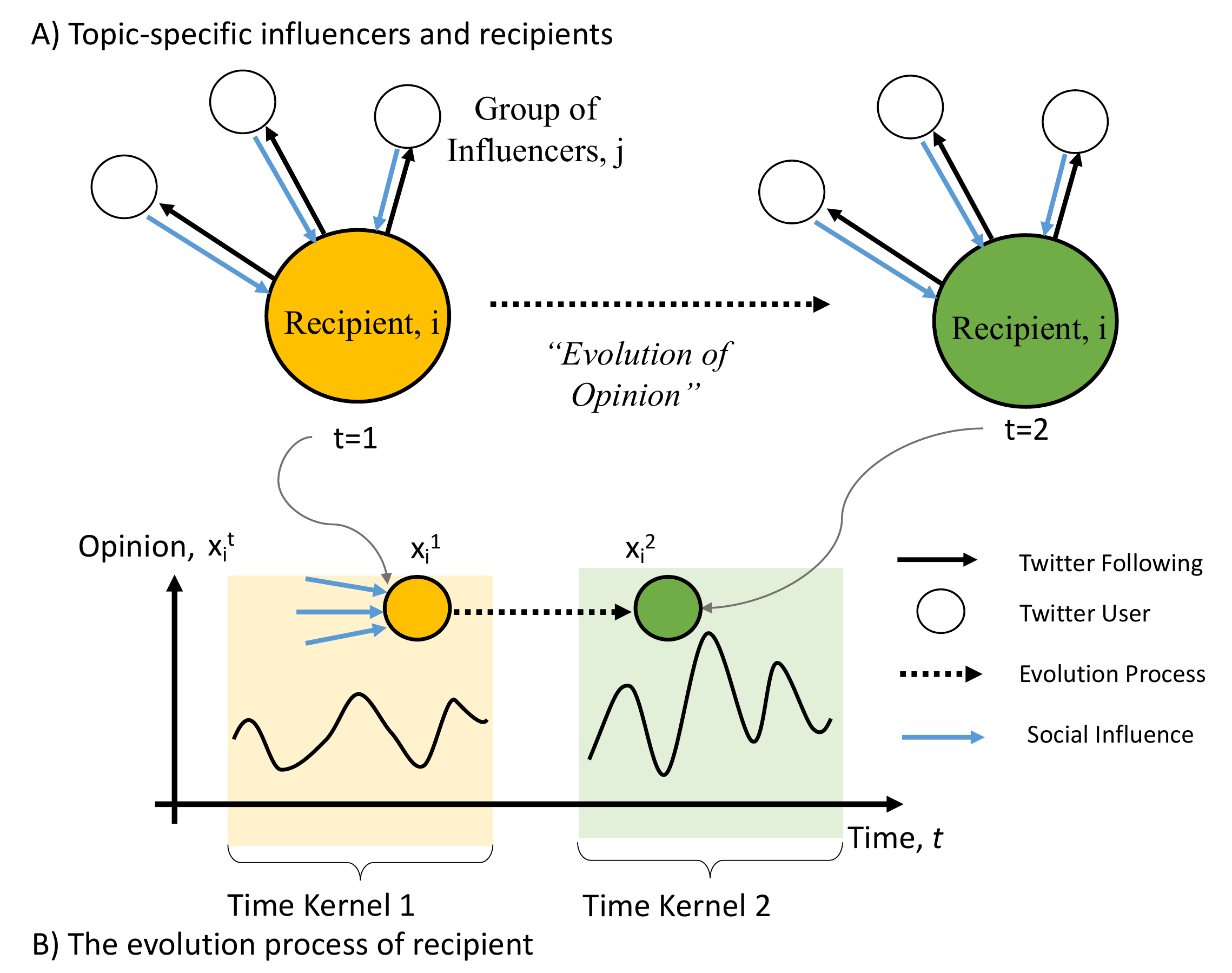}
  \caption{A) shows the recipient $i$ and the influencers $j$ (determined by Twitter following), and the influencers provide the forces of social influence on the recipient's opinion over time. B) shows the evolution process of recipient $i$ under the forces from influencers with a time kernel.}
  \label{system diagram}
\end{figure}

In this part, we show the reader we apply the social influence modelling method to online social networks. Fig. \ref{system diagram} presents the way we model the personal opinion evolution process under the effect of the online social influence as a function. 

In Fig. \ref{system diagram}A) we show two example types of Twitter users within a defined topic: left) the recipient $i$ and the influencers $j$ (determined by Twitter following), and right) over time, the influencers provide the forces of social influence on the recipient's opinion, leading to the evolution of its opinion. Our aim is modelling the evolution process, so a time kernel is applied to capture the modification of opinion as shown in Fig. \ref{system diagram}B). The size of the time kernel is selected to capture sufficient activity within a period (e.g., typically 10 days).

We use $x$ to represent the quantified opinion (compressed from aforementioned word embedding), and the opinion of recipient $i$ at time kernel $t$ is defined as $x_i^t$. We use function $g(x_i^1, x_j^1)$ represent the social influence from $j$ to $i$ at time $t=1$. However, the nature of $g(\cdot)$ is unknown at this time and has to be derived either from previous experimentation (see below) or function discovery. The linear combination of previous opinion with confidence weight $w_{ii}$ and social influence with influence weight $w_{ij}$ contributes to the opinion evolution model 
\begin{equation} \label{fun1}
x_i^2 = w_{ii}x_i^1 +\sum_{j, j\neq i}w_{ij}g(x_i^1, x_j^1)
\end{equation}

Using French's formal theory \cite{french1956formal}, in this paper, we model the discrepancy of opinions $x_i^t$ and $x_j^t$ to determine the effect from influencer $j$ to recipient $i$. So the influence effect is determined to be proportional to the size of the difference between their opinions $g(x_j^t, x_i^t) = (x_j^t-x_i^t)$. Beyond the function, there may include influence weights ($w_{ij}$) representing the strength of the effect. Formally, social pressure on the recipient $i$ is the sum of the effect from all influencers j conditioned by the weight ($w_{ij}$) of the tie between i and j $(-1.0 \leq w_{ij} \leq 1.0)$. The self-weight ($w_{ii}$) of the recipient $i$ represent to what degree the recipient is anchored on his previous position $(-1.0 \leq w_{ii} \leq 1.0)$ \cite{myers1982polarizing}. The influence process takes place gradually, as the influencer changes its position over time and influences the recipient toward its position. For each recipient, the discrete-time interpersonal influence mechanism can be describe as a ordinary differential equation

\begin{equation} \label{funi}
x_i^{t+1} = w_{ii}x_i^{t} +\sum_{j, j\neq i}w_{ij}(x_j^t-x_i^t)
\end{equation}

% where:
% \begin{description}
% \item[$x_i^t$] is opinion of user $i$ at time kernel $t$
% \item[$w_{ij}$] is influence weight from j to i
% \end{description}

Given the opinion evolution model, our approach is to: (1) identify topic-specific influencers and recipients using Twitter, (2) apply a time kernel to analyse opinion evolution, and (3) fit the data to empirical psychology ODE models and find the influencer weight.

\section{Case Study: Pipeline \& Data}

Here we choose COVID-19 as our specific topic. COVID-19 pandemic has been an ongoing global pandemic since December 2019. Discussions on disease symptoms, prevention, vaccine, and local policies widely spread online. From January 2020 to September 2021, over 35 million unique users post over 198 million Twitter using Covid-19 related keywords.
% \cite{gupta2020covid}. 
% Several studies analysed the evolution of public sentiment on COVID-19 to give a picture of how people experience under corresponding situations or policies\cite{lwin2022evolution}.
Our work concentrates on personal opinions evolution with the influence from the online network during the pandemic.

\subsection{Available Users}

\begin{figure}[t]
  \centering
  \includegraphics[width=3in]{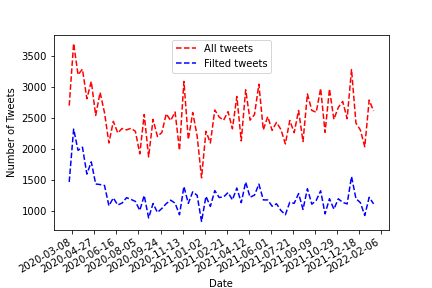}
  \caption{Number of all Tweets and topic-specific Tweets from March 2020 to Feb 2022}
  \label{NumTweets}
\end{figure}

Our model assumes that topic-specific influencers provide the forces of social influence on the recipient's opinion, leading to the evolution of opinion on the recipient. In the case study, we first need to locate these two types of Twitter users as the nodes in the opinion influence network: the recipient $i$ and the influencer $j$. We use the actual Twitter "Following" relationship to build the edges of the influence network, assuming that the recipient receives the forces of influence from their "Following" accounts.

To capture the opinion evolution process in a long time range, only the "active users" are under consideration for both recipient and influencer. In the research on the communication effect of mass media, the concept of "active users" is defined as users with a minimum level of activity.
% \cite{levy1984audience}. 
In our case, we set 10 days as one time period and determined the minimum standard as posting more than one topic-specific Twitter in at least 60 time periods from the 1st of March 2020 to the 30th of January 2022 (700 days). The missing periods would inherit the previous $t-1$ value of the opinion.

% \begin{table}[ht]
% \caption{15 COVID-19 Topic Keywords}
% \centering
% \begin{tabular}{| c c c|}
% \hline
%     omicron & coronavirus & koronavirus \\ covid & corona & isolation \\ quarantine & cdc & wuhancoronavirus \\ wuhanlockdown & ncov & wuhan \\ N95 & kungflu & epidemic \\[1ex]
% \hline
% \end{tabular}
% \label{table:keywords}
% \end{table}

To start building up the active user network, we first look into a list of COVID-19 experts on Twitter, including medical professionals, data scientists, and journalists. We used the Twitter API to gather each user's "Following" relationship and tweet contents. We manually pick a set of 15 keywords representing the COVID-19 topic. Then we filter the topic-specific tweets that contain at least one of the keywords. Fig.\ref{NumTweets} shows the number of all Tweets and topic-specific Tweets generated from active users from March 2020 to Feb 2022, including 175624 tweets and 85946 topic-related tweets in total.

Finally, we found 87 active recipients, and the mean value of the number of influencers per recipient is 17.655. Some recipients share some of the same influencers, and one recipient may act as an influencer in another recipient's network. 
% Fig.\ref{Connection} presents the whole user Connection Graph where each user's size is the value of its betweenness Centrality, with edges representing the "Following" relationship. 
Although the following links appear between the influencers, the interactions between influencers are not considered in the recipient's model.

% \begin{figure}[h]
%   \centering
%   \includegraphics[width=2.5in]{BetweenessCentrality.pdf}
%   \caption{User Connection Graph with size as Betweeness Centrality}
%   \label{Connection}
% \end{figure}

% \begin{figure}[ht]
%   \centering
%   \includegraphics[width=3.5in]{NumInfluencer.png}
%   \caption{Number of Influencers Distribution}
%   \label{NumInfluencer}
% \end{figure}

\subsection{Word Embedding and Compression to Uni-dimensional Opinion}
% \begin{figure*}[ht]
%   \centering
%   \includegraphics[width=5in]{Word_Embedding_A}
%   \caption{A) shows an example process of word-embedding and dimensionality.The similar relationship can be easily found by $v("man")-v("woman") \approx v("king")-v("queen")$. B) presents the visualization of 2-dimensional vectors of users' opinions on COVID-19 topic to give an impression of opinion evolution.}
%   \label{Word Embedding}
% \end{figure*}

\begin{figure}[t]
  \centering
  \includegraphics[width=3.45in]{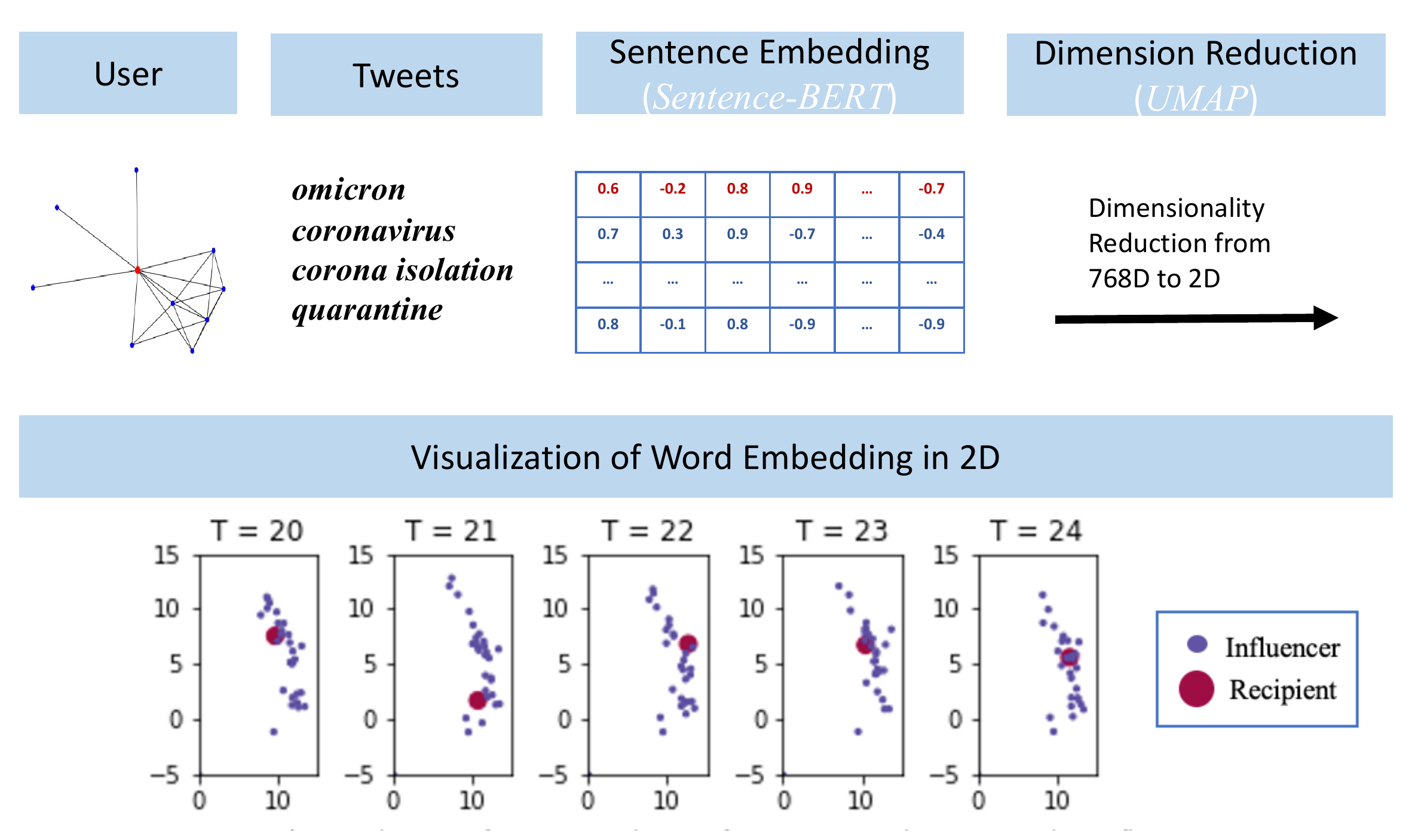}
  \caption{ Process of word-embedding, dimensionality and visualization of users' opinions on COVID-19 topic.}
  \label{Word Embedding}
\end{figure}
In this part, we present the reader the way we infer influence and opinion to regress our opinion model using online social media data.

The previously introduced social influence model mainly conceptualizes the opinion using pro-event and post-event psychology survey questions. In online social networks, we use the text content posted by one individual to represent the individual's opinion. We aim to apply the social influence modelling method to the online social network, so we use the compressed word-embedding vectors to capture the vibration of opinion. 
% Our model is based on the fact that group size, network structures, or relations among opinions determine the influence effect, and the opinion evolution mainly develops due to the impact.

% Then we aim to vectorize the tweet contents and compress the embedding representation to a uni-dimensional variable to fit the aforementioned opinion evolution models. Word-embedding method and dimensionality reduction ensure each vector's size, maintaining the syntactic and semantic meanings of words. 

% The example process of word-embedding and dimensionality is shown in Fig. \ref{Word Embedding}A). 7-dimensional word-embedding vectors are used to represent "man", "woman", "king", and "queen". With dimensional reduction projecting the 7-dimensional vectors to 2-dimensional vectors, the similar relationship can be easily found by $v("man")-v("woman") \approx v("king")-v("queen")$.

In our case study, we gather the COVID-19 specific tweets content as the initial input. We represent tweets using uni-dimensional continuum by word-embedding and dimensional reduction. 

The process is shown in Fig.\ref{Word Embedding}. For word-embedding, we use \textit{Sentence-BERT} with pretrained \textit{all-mpnet-base-v2} model \cite{reimers-2019-sentence-bert}. \textit{Sentence-BERT} takes sentences as the input data and produces 768-dimensional output vectors. 
% \textit{BERT} \cite{devlin2018bert} and \textit{RoBERTa} \cite{liu2019roberta} has shown the state-of-art performance on word-embedding based semantic textual similarity tasks.
\textit{Sentence-BERT} uses siamese and triplet structure on the pretrained \textit{BERT} network, leading to outperforming transfer learning tasks. 

We pass the topic-specific tweets from selected "active users" to \textit{Sentence-BERT} to obtain vector representations of each tweet, then take the average of the vectors in each time window. In this case, each user will have 70 768-dimensional word embedding vectors to represent their time-varying opinions on the COVID-19 topic.

Then each vector would be projected onto the uni-dimensional plane using the Uniform Manifold Approximation and Projection (UMAP) algorithm \cite{mcinnes2018umap}. UMAP is a scalable algorithm for dimension reduction, searching for a low-dimensional data projection with the closest topological structure. 
% The performance of UMAP is competitive with t-distributed Stochastic Neighbor Embedding\cite{van2008visualizing}  with no computational restrictions on embedding dimension. 
After UMAP compresses the 768-dimensional word-embedding vectors to 1-dimensional vectors, each user then has 70 uni-dimensional vectors representing their opinion evolution.

In Fig. \ref{Word Embedding}, we also present the visualization of 2-dimensional vectors of users' opinions on this topic to give an impression of opinion evolution. Red and blue dots represent the opinions from one of the recipients and the corresponding thirty influencers. The five sub-graphs visualize the variations of opinions from time kernel 20 to time kernel 25. It should be noted that the 2-dimensional vectors are only used in this visualization, and the opinion vectors would be compressed to uni-dimensional during model fitting to match the opinion evolution model.

\section{Opinion Model Evaluation}

\subsection{Multi-linear Regression}

% \begin{figure*}[ht]
%   \centering
%   \includegraphics[width=7in]{Regression.png}
%   \caption{A) shows the process of building multi-linear regression. The independent variables are $x_i^{t}$ and all corresponding $x_j^t$, where the dependant variable is $x_i^{t+1}$. B) 4 sub-graphs show the partial regression plots based on 4 independent variables $x_i^t$, $x_{j1}^t$, $x_{j2}^t$ and $x_{j3}^t$, including the regression lines with the observation dots.}
%   \label{Regression}
% \end{figure*}

Here we have the quantified opinions from 87 recipients and the corresponding influencers in 70 time kernels. We will then use the multi-linear regression method to evaluate the model performance in generating the opinion evolution process. 

For each recipient, we would build a regression process on the model 

\begin{equation} \label{funRegression}
x_i^{t+1} = w_{ii}x_i^{t} +\sum_{j, j\neq i}w_{ij}(x_j^t-x_i^t) = \beta_{ii}x_i^t+\sum_{j, j\neq i}\beta_{ij}x_j^t
\end{equation}

% The actual process is shown in Fig.\ref{Regression}A). 
The independent variables are $x_i^{t}$ and all corresponding $x_j^t$, where the dependant variable is $x_i^{t+1}$. In this case, all the number of observation is 69 since we capture 70 time windows. But the number of independent variables is varying depending on the number of influencers $n$. We use the ordinary least square (OLS) method to estimate the coefficients of multi-linear regression. The OLS method searching the coefficients by minimizing the sum of square errors between the observed and predicted values. Here, the coefficients $\beta_{ij}$ represent the influence weights of $w_{ij}$, and the self-weight $w_{ii}$ could be calculated from the coefficients $\beta_{ii}$ and $\beta_{ij}$. 
% Using the case of recipient7, Fig.\ref{Regression}B) shows the regression lines with the observation dots. The 4 sub-graphs show the partial regression plots based on 4 independent variables $x_i^t$, $x_{j1}^t$, $x_{j2}^t$ and $x_{j3}^t$.

% The following section will display several statistical measures on each regression model to show the goodness of fitting.

\subsection{Regression Result}

\begin{table}
  \caption{OLS Regression Result}
  \label{regression}
  \centering
  \begin{tabular}{ccl}
    \hline
    \hline
    No. of Influence Models & Observations per Model\\
    87 & 69 \\
    \hline
    $\hat{R}$ Mean & Adj. $\hat{R}$ Var \\0.98232 & 0.00769 \\
    Pro F-statistic Mean & Pro F-statistic Var\\
    0.00012 & 1.26e-06\\
    \hline
    \hline
\end{tabular}
\end{table}

The results of the regression models are shown in Table.\ref{regression}. 
We have 87 influence models and 69 observations per model. For each model, the number of influencers depends on different recipients, where the mean value of the number of influencers is 17.655, and the variance is 123.07.

The following two lines reveal the attributes that describe the performances of our multi-linear regressions, adjusted R-squared, and probability of F-statistic.

The adjusted R-squared score shows the explanatory power of regression models that contain multiple predictors. In 87 regression models, the mean adjusted R-squared score is 0.98232, and the variance is 0.00769, representing that the influencers can explain at least 0.98\% of the variance for the recipient's opinion. 

The null hypothesis of the F-statistic is that the effects of the predictors are 0. The F-statistic probability shows the probability of rejecting the null hypothesis, which indicates if the group of independent variables is essential. All probabilities of F-statistic in our models are close to zero with slight variance, representing the statistically significant of the predictors.

In summary, the high adjusted R-squared values and probabilities of the F-statistic reveal the remarkable explanatory power and statistical significance of the predictors.

\section{Conclusion \& Future Work}

This paper aims to model the personal opinion evolution
process under the effect of the online social influence as a
function. We propose the social network opinion ODE model, which considers both individual behaviour and network structure. We use Twitter empirical data to fit the parameters of the model. To achieve this, we introduced a pipeline to quantitatively represent personal opinion evolution with real-time Twitter data under the COVID-19 topic. Using the quantified real-time data as input, at least based on this topic, the opinion data indicate that social media users primarily shift their opinion based on influencers they follow and self-evolution of opinions over a long time scale is limited compared to the influences from others. 

Our next step is to analyze the relationships between the estimated influence weights and the actual interaction activities between users, revealing why and how influencers can be influential. We will also discover the social influence function through data-driven function discoveries allowing non-linear assumptions for diverse other topics.

\section*{Acknowledgment}
The work is supported by "Networked Social Influence and Acceptance in a New Age of Crises", funded by USAF OFSR under Grant No.: FA8655-20-1-7031.

\bibliographystyle{IEEEtran}
\bibliography{sample-base}

\begin{thebibliography}{10}
\providecommand{\url}[1]{#1}
\csname url@samestyle\endcsname
\providecommand{\newblock}{\relax}
\providecommand{\bibinfo}[2]{#2}
\providecommand{\BIBentrySTDinterwordspacing}{\spaceskip=0pt\relax}
\providecommand{\BIBentryALTinterwordstretchfactor}{4}
\providecommand{\BIBentryALTinterwordspacing}{\spaceskip=\fontdimen2\font plus
\BIBentryALTinterwordstretchfactor\fontdimen3\font minus
  \fontdimen4\font\relax}
\providecommand{\BIBforeignlanguage}[2]{{%
\expandafter\ifx\csname l@#1\endcsname\relax
\typeout{** WARNING: IEEEtran.bst: No hyphenation pattern has been}%
\typeout{** loaded for the language `#1'. Using the pattern for}%
\typeout{** the default language instead.}%
\else
\language=\csname l@#1\endcsname
\fi
#2}}
\providecommand{\BIBdecl}{\relax}
\BIBdecl

\bibitem{flanagin2017online}
A.~J. Flanagin, ``Online social influence and the convergence of mass and
  interpersonal communication,'' \emph{Human Communication Research}, vol.~43,
  no.~4, pp. 450--463, 2017.

\bibitem{conflict19}
N.~Tkachenko and W.~Guo, ``Conflict detection in linguistically diverse online
  social networks: a russia-ukraine case study,'' in \emph{ACM International
  Conference on Management of Digital EcoSystems (MEDES)}.\hskip 1em plus 0.5em
  minus 0.4em\relax ACM, 11 2019.

\bibitem{influencenetwork19}
D.~Centola, ``{Influential Networks},'' \emph{Nature Human Behaviour}, vol.~3,
  no.~7, 2019.

\bibitem{biran-etal-2012-detecting}
\BIBentryALTinterwordspacing
O.~Biran, S.~Rosenthal, J.~Andreas, K.~McKeown, and O.~Rambow, ``Detecting
  influencers in written online conversations,'' in \emph{Proceedings of the
  Second Workshop on Language in Social Media}.\hskip 1em plus 0.5em minus
  0.4em\relax Montr{\'e}al, Canada: Association for Computational Linguistics,
  Jun. 2012, pp. 37--45. [Online]. Available:
  \url{https://aclanthology.org/W12-2105}
\BIBentrySTDinterwordspacing

\bibitem{asch1956studies}
S.~E. Asch, ``Studies of independence and conformity: I. a minority of one
  against a unanimous majority.'' \emph{Psychological monographs: General and
  applied}, vol.~70, no.~9, p.~1, 1956.

\bibitem{bond201261}
R.~M. Bond, C.~J. Fariss, J.~J. Jones, A.~D. Kramer, C.~Marlow, J.~E. Settle,
  and J.~H. Fowler, ``A 61-million-person experiment in social influence and
  political mobilization,'' \emph{Nature}, vol. 489, no. 7415, pp. 295--298,
  2012.

\bibitem{french1956formal}
J.~R. French~Jr, ``A formal theory of social power.'' \emph{Psychological
  review}, vol.~63, no.~3, p. 181, 1956.

\bibitem{degroot1974reaching}
M.~H. DeGroot, ``Reaching a consensus,'' \emph{Journal of the American
  Statistical Association}, vol.~69, no. 345, pp. 118--121, 1974.

\bibitem{hegselmann2002opinion}
R.~Hegselmann, U.~Krause \emph{et~al.}, ``Opinion dynamics and bounded
  confidence models, analysis, and simulation,'' \emph{Journal of artificial
  societies and social simulation}, vol.~5, no.~3, 2002.

\bibitem{myers1982polarizing}
D.~G. Myers, ``Polarizing effects of social interaction,'' \emph{Group decision
  making}, vol. 125, pp. 137--138, 1982.

\bibitem{reimers-2019-sentence-bert}
\BIBentryALTinterwordspacing
N.~Reimers and I.~Gurevych, ``Sentence-bert: Sentence embeddings using siamese
  bert-networks,'' in \emph{Proceedings of the 2019 Conference on Empirical
  Methods in Natural Language Processing}.\hskip 1em plus 0.5em minus
  0.4em\relax Association for Computational Linguistics, 11 2019. [Online].
  Available: \url{https://arxiv.org/abs/1908.10084}
\BIBentrySTDinterwordspacing

\bibitem{mcinnes2018umap}
L.~McInnes, J.~Healy, and J.~Melville, ``Umap: Uniform manifold approximation
  and projection for dimension reduction,'' \emph{arXiv preprint
  arXiv:1802.03426}, 2018.

\end{thebibliography}

\end{document}